\documentclass[AMA,STIX1COL]{WileyNJD-v2}
\articletype{Research Article}%

\received{26 Aug 2019}
\revised{6 June 2016}
\accepted{6 June 2016}

\usepackage{epstopdf}
\usepackage{multirow} 
\usepackage{tikz}
\begin{document}

\title{Causal inference with recurrent data via inverse probability treatment weighting method (IPTW)}%\protect\thanks{This is an example for title footnote.}

\author[1]{Haodi Liang}

\author[2]{Cecilia Cotton}

\authormark{AUTHOR ONE \textsc{et al}}

\address[1]{\orgdiv{Department of Statistics and Actuarial Science}, \orgname{University of Waterloo}, \orgaddress{\state{Ontario}, \country{Canada}}}

\address[2]{\orgdiv{Department of Statistics and Actuarial Science}, \orgname{University of Waterloo}, \orgaddress{\state{Ontario}, \country{Canada}}}

\corres{Correspondence to: Haodi Liang. Tel: 226-989-3910 
\newline{}\ 
\email{h28liang@uwaterloo.ca}}

\presentaddress{200 University Ave W, Waterloo, ON N2L 3G1}

\abstract[Summary]{Propensity score methods are increasingly being used to reduce estimation bias of treatment effects for observational studies. Previous research has shown that propensity score methods consistently estimate the marginal hazard ratio for time to event data. However, recurrent data frequently arise in the biomedical literature and there is a paucity of research into the use of propensity score methods when data are recurrent in nature. The objective of this paper is to extend the existing propensity score methods to recurrent data setting. We illustrate our methods through a series of Monte Carlo simulations. The simulation results indicate that without the presence of censoring, the IPTW estimators allow us to consistently estimate the marginal hazard ratio for each event. Under administrative censoring regime, the stabilized IPTW estimator yields biased estimate of the marginal hazard ratio, and the degree of bias depends on the proportion of subjects being censored. For variance estimation, the naïve variance estimator often tends to substantially underestimate the variance of the IPTW estimator, while the robust variance estimator significantly reduces the estimation bias of the variance.}


%\jnlcitation{\cname{%
%\author{Williams K.}, 
%\author{B. Hoskins}, 
%\author{R. Lee}, 
%\author{G. Masato}, and 
%\author{T. Woollings}} (\cyear{2016}), 
%\ctitle{A regime analysis of Atlantic winter jet variability applied to evaluate HadGEM3-GC2}, \cjournal{Q.J.R. Meteorol. Soc.}, \cvol{2017;00:1--6}.}

\maketitle

%\footnotetext{\textbf{Abbreviations:} ANA, anti-nuclear antibodies; APC, antigen-presenting cells; IRF, interferon regulatory factor}

\section{Introduction}\label{sec1}
Causal inference is an emerging field in statistics. In medical research, we are often interested in understanding the effect of treatment on an outcome. The gold standard is to conduct an experimental study where treatment is randomized. However, observational studies frequently arise in medical research, in which treatment assignment is often related to characteristics of patients. As a result, the characteristics of treatment and control group may be systematically different. To this end, adjustments must be made to balance the covariates between the two groups. Over the past several decades, different methods for making causal inference with observational data have been developed and there has been an increasing interest in using the propensity score methods. The propensity score is defined as the probability of treatment assignment conditional on measured baseline covariates \cite{Rubin1}. There are four propensity score methods that are used most often in the biomedical literature: matching, stratification, inverse probability weighting and covariate adjustment. These methods allow us to reconstruct a pseudo-sample which reflects an experimental data setting, thus reducing or eliminating bias in estimating the treatment effect.

In the 1980's, researchers mainly focused on bias reduction on a linear scale. Rosenbaum and Rubin (1983) demonstrated that by dividing the sample into five mutually exclusive equal-sized strata based on the propensity score would result in an over 90\% bias reduction when the treatment is an average treatment effect (ATE) \cite{Rubin1,Rubin3}. In addition to that, the inverse probability treatment weighting method using the propensity score was proven to unbiasedly estimate the average treatment effect (ATE) \cite{Rubin1}. It was not until recent years that propensity score methods for non-linear measures of treatment effects received attention. Some applied researchers tried to use propensity score methods to estimate non-linear treatment effects such as odds ratio and hazard ratio. However, the degree of bias incurred had not been extensively studied \cite{austin1}. It was not until the beginning of the $21^{st}$ century that Austin (2007) performed a series of Monte Carlo simulation studies to examine the degree of bias when treatment effects are measured using a hazard ratio, odds ratio and rate ratio \cite{austin1}. The simulation results indicated that conditional on the propensity score, matching, stratification and inverse probability weighting all resulted in biased estimates of the true conditional hazard ratio and odds ratio. Interestingly the rate ratio was consistently estimated for all propensity score methods. This is because conditional on the propensity score, we estimate the marginal treatment effect instead of the conditional treatment effect \cite{Rosenbaum}. A conditional effect refers to the average effect at the individual level, of removing a subject from treated to untreated. The regression coefficient of treatment indicator in a Cox proportional hazards model is a conditional effect \cite{austin2}. A marginal effect is the average effect at the population level, of moving the whole population from treated to untreated \cite{Greenland}. For randomized controlled trials, we estimate the marginal effect by using the Cox proportional hazard model with treatment indicator as the only covariate. Austin (2007) concluded that the marginal treatment effect coincides with the conditional treatment effect when the measure of treatment effect is a difference in means or rate ratio while they do not coincide when the measure of treatment effect is odds ratio or hazard ratio \cite{austin1} \cite{austin4}. 

Many observational data in real life are recurrent in nature. We already know that the propensity score methods estimate the marginal hazard ratio for time to event data \cite{austin2}. However, there is a paucity of research on making causal inference for recurrent data. Hence it would be desirable to extend the propensity score framework to two events an possibly multiple events. Therefore, the objective of this paper is to formulate appropriate propensity score methods to estimate the treatment effect in the context of recurrent data.

\section{Methods}\label{sec2}
In this section we investigate the inverse probability of treatment weighting method (IPTW) for estimating treatment effects when there are two events for each subject. Three scenarios are discussed in this section: independent gap times, time-varying covariates with fixed treatment and time-varying covariates with time-varying treatment \cite{Cook}. For each scenario, model assumptions and specifications of the treatment model and the outcome model are given.

\subsection{Notation and Model Setup}
We use the following notation throughout this chapter. Assume there are $n$ subjects $i = 1, 2, ..., n$ although we suppress $i$ notation in this chapter. Let $X(j)$ be a $p$-dimensional vector of covariates at the start of the $j^{th}$ gap time and $Z(j)$ be treatment status at the start of the $j^{th}$ gap time. Let $W_{1}$ denote the first gap time and $W_{2}$ denote the second gap time. We define the propensity score for the first event $e_{1}$ to be 
$$e_{1} = P(Z(1)=1|X(1)=x(1))$$
We define $e_{2}$ to be the propensity score for the second gap time i.e. the probability of treatment for the second event conditional on all past covariate and treatment history. That is, 
$$e_{2} = P(Z(2)=1|\overline{X}(2)= \overline{x}(2), \overline{Z}(1)= \overline{z}(1))$$
where $\overline{X}(j) = (X(1), X(2), ... X(j))$ is the covariate history through the start of the $j^{th}$ gap time and $\overline{Z}(j) = (Z(1), Z(2), ... Z(j))$ is the treatment history through the start of the $j^{th}$ gap time. Hence, the probability of a subject receive treatment for both events conditional on all past history is:
\begin{align*}
e_{1}e_{2} 
&= P(Z(1)=1,Z(2)=1|\overline{X}(2) =\overline{x}(2), \overline{Z}(1) = \overline{z}(1))
\\&= E(Z(1)Z(2)|\overline{X}(2) =\overline{x}(2), \overline{Z}(1) = \overline{z}(1))
\end{align*}

Intuitively, the IPTW weights are defined to be the inverse of the probability of treatment path conditional on all past treatment and covariate history. The stabilized inverse probability treatment weights are \cite{Cole}
$$sw_{1} = \frac{P(Z(1)=z(1))}{P(Z(1)=z(1)|X(1)=x(1))} = \frac{P(Z(1)=1)Z(1)}{{e}_{1}} + \frac{P(Z(1)=0)(1-Z(1))}{1-e_{1}}$$
and the stabilized IPTW weights for the second event are
\begin{align*}
sw_{2} &= \frac{P(Z(1)=z(1))}{P(Z(1)=z(1),X(1)=x(1))} \cdot \frac{P(Z(2)=z(2)|Z(1)=z(1))}{P(Z(2)=z(2)|\overline{X}(2)=\overline{x}(2),\overline{Z}(1)=\overline{z}(1))}
\\&= p_{11} \frac{Z(1)Z(2)}{e_{1}e_{2}} + p_{10}\frac{Z(1)(1-Z(2))}{e_{1}(1-e_{2})} + p_{01}\frac{(1-Z(1))Z(2)}{(1-e_{1})e_{2}} + p_{00} \frac{(1-Z(1))(1-Z(2))}{(1-e_{1})(1-e_{2})}
\end{align*}
where $p_{ij} = P(Z(1)=i, Z(2)=j)$ \cite{Hernan}\cite{Brumback}.

The reason why we consider the stabilized weights is that the conventional weights sometimes result in extremely large weights for a few subjects. As a result, these subjects dominate the weighted analysis, and this results in unstable estimation of the marginal hazard ratio. In addition, the use of the conventional weights sometimes also leads to rather large variance for the conventional IPTW estimator \cite{Hernan}. 

\subsection{Time-Fixed Treatment and Covariates}
We start with the simplest case where there are two independent gap times $W_{1}$ and $W_{2}$ for each subject. For simplicity we assume $X$ is a 1-dimensional scalar. Figure 1 \cite{Pearl}  illustrates the relationship among $X$, $Z$, $W_{1}$ and $W_{2}$. Our goal is to use propensity score methods to consistently estimate the marginal treatment effect. In this setting we assume treatment and covariates are fixed over time, so we use $Z$ and $X$ without the $j$ notation. Define $e$ to be the probability of treatment conditional on covariates. We regress treatment indicator $Z$ on $X$ to obtain the estimated propensity score $\hat{e}$:
$$ \hat{e} = expit\big(\hat{\alpha}_{0} + \hat{\alpha}_{1}x\big)$$
To estimate the overall marginal treatment effect, we regress the survival outcomes $W_{1}$ and $W_{2}$ on the treatment status $Z$ through a weighted Cox proportional hazards model with the stabilized weights as defined in Section 2.1:
$$ h_{j}(w|x,z) = h_{0}(w)e^{\beta^{m}z}$$
where $j$ = 1, 2.

\begin{figure}[h]
\begin{center}
\begin{tikzpicture}[scale=1]
\node (Z1) at (-1.5,1.5) {$Z$};
\node (W1) at (0,0) {$W_{1}$};
\node (W2) at (2,0) {$W_{2}$};
\node (X1) at (-1.5,-1.5) {$X$};

\draw[->] (X1) -- (Z1);
\draw[->] (X1) -- (W1);
\draw[->] (Z1) -- (W1);
\draw[->] (X1) --(W2);
\draw[->] (Z1) -- (W2);

\end{tikzpicture}
\caption{Causal graph for time-fixed treatment and covariate}
\end{center}
\end{figure}
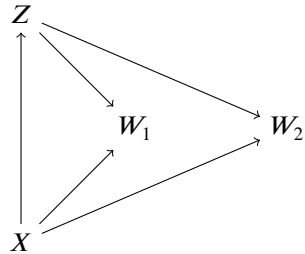

We perform a simulation study to examine the numerical performance of the IPTW estimator using the stabilized weights. For simplicity, we generate a standard normal covariate $X$. For each subject we generate a treatment probability through a logistic regression model:
$$\pi = expit\big(\alpha_{0} + \alpha_{1}x\big)$$
Then we generate a treatment status for each individual $Z \sim Bernoulli(\pi)$.  We set $\alpha_{0}$ to be -1.1392 by a bisection approach to achieve an overall treatment prevalence of 25\% \cite{austin2}. Here $\alpha_{1}$ represents the log odds ratio of receiving treatment per unit increase in $X$ and we set it to be log(1.5). For each subject we simulate two independent gap times $W_{1}$ and $W_{2}$ from a Cox proportional hazards model. We choose the baseline hazard to be an exponential distribution with $\lambda$ = 1. Hence, the hazard takes the form:
$$h_{j}(w|x,z) = e^{\beta^{c}z+\beta_{1}x}$$
where $j$ =1 ,2. The association parameter between $X$ and $W_{j}$ is $\beta_{1}$, and is set to log(1.5). The simulation algorithm for $W_{1}$ and $W_{2}$ is as follows \cite{Bender}

\begin{itemize}
\item Simulate two independent standard uniform distribution $u_{1}$ and $u_{2}$
\item Simulate $w_{1} = \dfrac{-log(u_{1})}{e^{\beta^{c}z + \beta_{1}x}}$ and
$w_{2} = \dfrac{-log(u_{2})}{e^{\beta^{c}z + \beta_{1}x}}$
\end{itemize}

The true marginal hazard ratio is determined as follows: For each subject we simulate both potential outcomes for both gap times under treatment and control conditions. Then we regress the gap time on treatment status to obtain the log of the true marginal hazard ratio $\beta^{m}$. The above data generation method is based on a conditional hazard ratio $e^{\beta^{c}}$. However, the IPTW estimator estimates the marginal hazard ratio. To this end, we use a bisection approach to determine $\beta^{c}$ that induces the specified marginal hazard ratio $e^{\beta^{m}}$ \cite{austin2}. 

To estimate $\beta^{m}$ for a given simulated dataset, first we obtain the estimated propensity score $\hat{e}$ through a logistic regression model. Then we calculate the stabilized weights $sw = \frac{P(Z=1)Z}{e} + \frac{P(Z=0)(1-Z)}{1-e}$. Finally we regress the gap times on the treatment indicator through a weighted Cox proportional hazards model:
$$ h_{j}(w|x,z) = h_{0}(w)e^{\beta^{m}z}$$
Doing this allows us to estimate the marginal treatment effect. Since weighting artificially creates a cluster for each subject, inducing a within-subject correlation, the naïve variance estimator often fails to correctly estimate the variance of $\hat{\beta}^{m}$ \cite{Hernan}. To address this issue, we use the robust variance estimator proposed by Lin \cite{Lin} \cite{Lin2}.

\subsection{Time-Varying Covariates}
Next we consider the case where covariates change over time while treatment is fixed. Assume $X(j)$ is a 1-dimensional scalar. The relationship among these variables is illustrated in Figure 2. Due to the presence of time-varying covariates, the marginal hazard ratio may differ for the two events. The methodology for estimating the marginal hazard ratio is as follows: First we estimate the propensity score through a logistic regression model:
$$\hat{e} = expit\big(\hat{\alpha}_{0} + \hat{\alpha}_{1}x(1)\big)$$
Then to estimate the marginal hazard ratio for the first and second event, we run a weighted Cox proportional hazards model using the stabilized weights, whose hazard takes the form:
$$ h_{j}(w|\overline{x}(j),\overline{z}(j)) = h_{0_{j}}(w)e^{\beta^{m_{j}}z}$$
where $j$ = 1, 2.

\begin{figure}[h]
\begin{center}
\begin{tikzpicture}[scale=1]
\node (Z1) at (-1.5,1.5) {$Z(1)$};
\node (W1) at (0,0) {$W_{1}$};
\node (W2) at (5,0) {$W_{2}$};
\node (X1) at (-1.5,-1.5) {$X(1)$};
\node (X2) at (3.5,-1.5) {$X(2)$};

\draw[->] (X1) -- (Z1);
\draw[->] (X1) -- (W1);
\draw[->] (Z1) -- (W1);
\draw[->] (Z1) -- (W2);
\draw[->] (X2) --(W2);
\draw[->] (X1) -- (X2);
\end{tikzpicture}
\caption{Causal graph for time-varying covariates}
\end{center}
\end{figure}
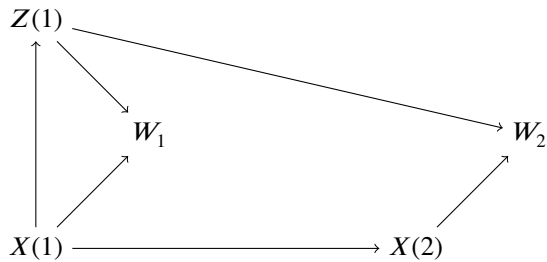

We perform the following simulation study to examine the numeric performance of the proposed IPTW estimator. We follow the same data generation methods described in Section 2.2 for the first gap time. Based on the first gap time $W_{1}$ and the covariate $X(1)$, we simulate the $X(2)$ covariate a second dependent gap time $W_{2}$ as follows \cite{Bender}:
\begin{itemize}
\item Simulate a standard uniform random variable $u_{2}$
\item Simulate a random variable $v \sim N(0,16)$ independent of $x(1)$ and $u_{2}$ 
\item Set $x(2) = x(1) + v$
\item Simulate $w_{2} = \dfrac{-log(u_{2})}{e^{\beta^{c}z+\beta_{1}x(2)}}$
\end{itemize}

Although the above data generation for the second gap time results in the same conditional hazard ratio $e^{\beta^{c}}$ for both gap times, the marginal hazard ratio may differ. For a given conditional hazard ratio $e^{\beta^{c}}$, we determine the true marginal hazard ratio for the second event using a similar method to that discussed in Section 2.2. 

We obtain the estimated propensity score $\hat{e}$, along with the stabilized weights for both gap times. Then, we regress the gap time $W_{j}$ on treatment indicator $Z(j)$ through a weighted Cox proportional hazards model with the stabilized weights $sw_{1}$ and $sw_{2}$ to estimate the marginal hazard ratio for the first and second event. 
$$h_{j}(w|\overline{x}(j),z) = h_{0_{j}}(w)e^{\beta^{m_{j}}z}$$
where $j$ = 1, 2. Here $\beta^{m_{j}}$ denotes the log marginal hazard ratio for the $j^{th}$ gap time. Finally, we estimate the variance of $\hat{\beta}^{m_{1}}$ and $\hat{\beta}^{m_{2}}$ using both the naïve variance estimator and the robust variance estimator.

\subsection{Time-Varying Treatment and Covariates}
We make further extensions by considering both time-varying treatment and covariates. Assume $X(j)$ is a 1-dimensional scalar. In such a setting, treatment status of the second event $Z(2)$ is dependent on treatment status of the first event and covariate value at the start of the second gap time $X(2)$. Figure 3 illustrates the relationship among these variables. In this setting due to time-varying treatment, for each subject we need to estimate two propensity scores. This can be done through the following logistic regression models:
$$ \hat{e}_{1} = expit\big(\hat{\alpha}_{0} + \hat{\alpha}_{1}x(1)\big)$$ 
$$ \hat{e}_{2} = expit\big(\hat{\gamma}_{0} + \hat{\gamma}_{1}x(2) + \hat{\gamma}_{2}z(1)\big)$$
Then we estimate the marginal hazard ratio for the first and second gap time through the following weighted Cox proportional hazards model using the stabilized weights $sw_{1}$ and $sw_{2}$:
$$ h_{j}(w|\overline{x}(j),z(j)) = h_{0_{j}}(w)e^{\beta^{m_{j}}z(j)}$$

We illustrate our methodology for estimating the marginal treatment effect through a simulation study. We use the previously discussed data generation methods for the first gap time. We consider the following dependence relationship between $X(1)$ and $X(2)$: 

\begin{figure}[h]
\begin{center}
\begin{tikzpicture}[scale=1]
\node (Z1) at (-1.5,1.5) {$Z(1)$};
\node (Z2) at (3.5,1.5) {$Z(2)$};
\node (W1) at (0,0) {$W_{1}$};
\node (W2) at (5,0) {$W_{2}$};
\node (X1) at (-1.5,-1.5) {$X(1)$};
\node (X2) at (3.5,-1.5) {$X(2)$};

\draw[->] (X1) -- (Z1);
\draw[->] (X1) -- (W1);
\draw[->] (Z1) -- (W1);
\draw[->] (X2) --(Z2);
\draw[->] (X2) --(W2);
\draw[->] (Z2) -- (W2);
\draw[->] (Z1) -- (Z2);
\draw[->] (X1) -- (X2);
\end{tikzpicture}
\caption{Causal graph for time-varying treatment and covariate}
\end{center}
\end{figure}
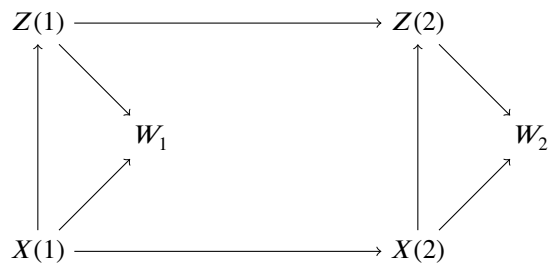

$$X(2) = X(1) + Y $$ where $Y \perp X(1)$ and $Y \sim N(0,16)$.

The above data generation technique results in a correlation of 0.24 between $X(1)$ and $X(2)$. Then, we simulate treatment status for the second gap time $Z(2)$ as follows. First we simulate a treatment probability through a logistic regression model:
$$logit(\pi_{2}) = \gamma_{0} + \gamma_{1}x(2) + \gamma_{2}z(1)$$
We set $\gamma_{1}$ to log(1.5) and $\gamma_{2}$ to log(1.5). Hence the log odds ratio of receiving treatment when t = 2 is 1.5 per one unit increase in $X(2)$ keeping treatment at t = 1 the same. We allow treatment prevalence for the second event to be 25\% or 50\%. We set $\gamma_{0}$ to be -1.7233 to achieve an overall treatment prevalence of 25\% and -0.1000 to achieve an overall treatment pravalence of 50\% for the second event \cite{austin3}. Having set all the parameters for the treatment model, we generate treatment status $Z(2) \sim Bernoulli(\pi_{2})$. We simulate the first and second gap time $W_{1}$ and $W_{2}$ from a Cox proportional hazards model, whose hazard takes the form:
$$h_{j}(w|\overline{x}(j),\overline{z}(j)) = h_{0}(w)e^{\beta^{c}z(j)+\beta_{1}x(j)}$$
 We follow the similar method to that described in Section 2.2 to obtain the true marginal hazard ratio for both gap times. 

For a given simulated dataset, to estimate the marginal hazard ratio for both gap times, first we obtain the estimated propensity score $\hat{e}_{1}$ and $\hat{e}_{2}$ through the following logistic regression models:
$$\hat{e}_{1} = expit\big(\hat{\alpha_{0}} + \hat{\alpha_{1}}x(1)\big)$$ and
$$\hat{e}_{2} = expit\big(\hat{\gamma_{0}}+ \hat{\gamma_{1}}x(2) + \hat{\gamma_{2}}z(1)\big)$$
Then we regress the first gap time on treatment indicator $Z(1)$ through a Cox proportional hazards model using the stabilized weights $sw_{1}$ from Section 2.1 to obtain the estimated marginal hazard ratio $e^{\hat{\beta}^{m_{1}}}$ for the first gap time. We run another weighted Cox proportional hazards model to regress the second gap time on treatment indicator $Z(2)$ using the stabilized weights $sw_{2}$ to obtain the estimated marginal hazard ratio $e^{\hat{\beta}^{m_{2}}}$ for the second gap time. The hazard takes the form:
$$h_{j}(w|\overline{x}(j),z(j)) = h_{0_{j}}(w)e^{\beta^{m_{j}}z(j)}$$
where $j$ = 1, 2. Finally, we estimate the variance of $\hat{\beta}^{m_{1}}$ and $\hat{\beta}^{m_{2}}$ using both the naïve variance estimator and the robust variance estimator.

\subsection{Administrative Censoring}
Often we have to deal with censored recurrent data where each subject experiences a different number of recurrent events. When censoring is a time-dependent confounder, previous methods for estimating the marginal treatment effect without adjustments for censoring may yield biased results. To this end we incorporate weights for censoring and we examine the performance of the IPTW estimator in the presence of censoring. We incorporate an administrative censoring time $\tau$. We define the censoring indicator $\delta_{1} = I(w_{1} \leq \tau)$ and $\delta_{2} = I(w_{1}+w_{2} \leq \tau)$. We can treat $(Z(i),\delta_{i})$ as a treatment vector at the start of the $i^{th}$ gap time. Thus, intuitively the IPTW weights are the inverse of the probability of treatment path of the subject.

The stabilized censoring weights are \cite{Hernan}
$$sw_{1}^{\dagger} = \frac{P(\delta_{1}=1)}{P(\delta_{1}=1|x(1),z(1))}$$
$$sw_{2}^{\dagger} = \frac{P(\delta_{1}=1)}{P(\delta_{1}=1|x(1),z(1))} \cdot \frac{P(\delta_{2}=1|\delta_{1}=1)}{P(\delta_{2}=1|\delta_{1}=1,\overline{x}(2),\overline{z}(2))}$$

We allow the administrative censoring time to be $\tau$ = 1 or 0.25.  When $\tau = 1$,  approximately 30\% of the subjects are censored for the first event and approximately 60\% of the subjects are censored for the second event. When $\tau = 0.25$, approximately 70\% of the subjects are censored for the first event and approximately 90\% of the subjects are censored for the second event. We use the stabilized weights defined above to estimate the marginal hazard ratio and its standard errors for the second gap time. The simulation results are available in Table 5-6.

\section{Simulation Results}
We allow the true marginal hazard ratio for the first gap time $e^{\beta^{m_{1}}}$ to be 1, 1.5, 2, 2.5 and 3. We determine the corresponding $\beta^{c}$ that results in the specified marginal hazard ratio using a bisection approach \cite{austin2}. For a given $\beta^{c}$, there is also a corresponding true marginal hazard ratio for the second gap time $e^{\beta^{m_{2}}}$. We summarize the relationship in Table 1.

For each of the three simulation settings, we simulate 1,000 datasets, each consisting of 10,000 subjects. In each of the 1,000 simulated datasets, we record the estimated log marginal hazard ratio for both gap times $\hat{\beta}_{1}(j)$ and $\hat{\beta}_{2}(j)$, along with its naïve standard error $\hat{\sigma}_{1}(j)$ and $\hat{\sigma}_{2}(j)$. We record the average estimated log marginal hazard ratio $\overline{\hat{\beta}}^{m_{k}} =\frac{1}{1,000} \sum_{j=1}^{1,000}\hat{\beta}_{k}(j)$ for $k$ = 1, 2. We define the average bias of the log marginal hazard ratio as:
$\frac{\overline{\hat{\beta}}^{m_{j}} - \beta^{m_{j}}}{\beta^{m_{j}}} \cdot100\%$ where $j$ = 1, 2. Then we determine the average standard error of the log hazard ratio across the 1,000 datasets: $ASE_{k} = \hat{\sigma}_{k} = \frac{1}{1,000}\sum_{j=1}^{1,000}\hat{\sigma}_{k}(j)$ where $k$ = 1, 2. We also determine the empirical standard error of the 1,000 estimated log marginal hazard ratios for both gap times:
 $ESE_{k} = \sqrt{\frac{\sum_{j=1}^{1,000}\big(\hat{\beta}_{k}(j)-\beta^{m_{k}}\big)^{2}}{1,000-1}}$ where $k$ = 1, 2 \cite{austin3}. If the variance of $\hat{\beta}^{m_{1}}$ and $\hat{\beta}^{m_{2}}$ are correctly estimated, the average standard error should be close to the empirical standard error. For each of the three simulation settings, we record the average estimated log marginal hazard ratio,  along with its average bias, average naïve standard error, average robust standard error and empirical standard error. We summarize the simulation results for the second gap time in tables below. We repeat the same procedures for a sample size of 500 and due to limited space, we move the results for scenarios with sample size of 500 to web material.

\begin{center}
\begin{table}[t]
\centering
\caption{Marginal and conditional log hazard ratios used in simulation study}
\begin{tabular*}{500pt}{@{\extracolsep\fill}lcccc@{\extracolsep\fill}}
\toprule
True log  & True marginal  & True log & True log& True marginal \\
marginal HR $\beta^{m_{1}}$  & HR $e^{\beta^{m_{1}}}$  & conditional HR $\beta^{c}$ & marginal HR $\beta^{m_{2}}$ & HR $e^{\beta^{m_{2}}}$ \\
\midrule
0 & 1 & 0 & 0 & 1 \\
0.4055 & 1.5 & 0.4599 & 0.2085 & 1.2318 \\
0.6931 & 2 & 0.7830 & 0.3551 & 1.4263 \\
0.9163 & 2.5 & 1.0313 & 0.4686 & 1.5978 \\
1.0986 & 3 & 1.2331 & 0.5616 & 1.7535\\
\bottomrule
\end{tabular*}
\begin{tablenotes} \scriptsize
\item[$\dagger$] HR: Hazard ratio
\end{tablenotes}
\end{table}
\end{center}

\begin{center}
\begin{table}[t]
\centering
\caption{Simulation results for independent gap times, sample size = 10,000}
\begin{tabular*}{500pt}{@{\extracolsep\fill}lccccccc@{\extracolsep\fill}}
\toprule
Prevalence = 25\%:\\
\toprule
True log  &  True marginal & Estimated log & Estimated & Avg & & & \\ 
 marginal  HR $\beta^{m_{2}}$& HR $e^{\beta^{m_{2}}}$  &  marginal HR $\overline{\hat{\beta}}^{m_{2}}$  &marginal HR $e^{\overline{\hat{\beta}}^{m_{2}}}$ & Bias &ASE &  ESE  & RSE \\
 \midrule
 0 & 1 & 0.0002 & 1.0002 & -0.02\% & 0.0163 & 0.0175 & 0.0196 \\
 0.4055 & 1.5 & 0.4091 & 1.5055 & 0.89\% & 0.0165 & 0.0191 & 0.0207 \\
 0.6931 & 2 & 0.7009 & 2.0156 & 1.13\% & 0.0168 & 0.0200 & 0.0216\\
 0.9163 & 2.5 & 0.9278 & 2.5289 & 1.26\% & 0.0172 & 0.0209 & 0.0223\\
 1.0986 & 3 & 1.1149 & 3.0493 & 1.48\% & 0.0175 & 0.0206 & 0.0230\\
\bottomrule
Prevalence = 50\%:\\
\toprule
 0 & 1 & 0.0000 & 1.0000 & 0.00\% & 0.0141 & 0.0140 & 0.0156 \\
 0.4055 & 1.5 & 0.4050 & 1.4993 & -0.12\% & 0.0144& 0.0143 & 0.0164 \\
 0.6931 & 2 & 0.6927 & 1.9991 & -0.06\% & 0.0148 & 0.0153 & 0.0174\\
 0.9163 & 2.5 & 0.9172 & 2.5023 & 0.10\% & 0.0153 & 0.0165 & 0.0182\\
 1.0986 & 3 & 1.0989 & 3.0009 & 0.03\% & 0.0157 & 0.0166 & 0.0190\\
\bottomrule
\end{tabular*}
\begin{tablenotes}\scriptsize
\item[$\dagger$ HR]: Hazard ratio
\item[$\ddagger$ ASE]: Average standard error
\item[$\dagger\dagger$ ESE]: Empirical standard error
\item[$\dagger\ddagger$ RSE]: Average robust standard error
\end{tablenotes} 
\end{table}
\end{center}

% Scenario 2 ------------------

\begin{center}
\begin{table}[t]
\centering
\caption{Simulation results for time-varying covariates, sample size = 10,000}
\begin{tabular*}{500pt}{@{\extracolsep\fill}lccccccc@{\extracolsep\fill}}
\toprule
Prevalence = 25\%: \\
\toprule
True log  &  True marginal & Estimated log & Estimated & Avg & & & \\ 
 marginal  HR $\beta^{m_{2}}$& HR $e^{\beta^{m_{2}}}$  &  marginal HR $\overline{\hat{\beta}}^{m_{2}}$  &marginal HR $e^{\overline{\hat{\beta}}^{m_{2}}}$ & Bias &ASE &  ESE  & RSE \\
 \midrule
 0 & 1 & -0.0004 & 0.9996 & -0.04\% & 0.0231 & 0.0244 & 0.0247 \\
0.2085 & 1.2318 & 0.2104 & 1.2342 & 0.91\% & 0.0232 & 0.0255 & 0.0257 \\
 0.3551 & 1.4263 & 0.3619 & 1.4360 & 1.91\% & 0.0232 & 0.0265 & 0.0266 \\
 0.4686 & 1.5978 & 0.4803 & 1.6166 & 2.50\% & 0.0233 & 0.0276 & 0.0274 \\
 0.5616 & 1.7535 & 0.5797 & 1.7855 & 3.22\% & 0.0235 & 0.0270 & 0.0281 \\
\bottomrule
Prevalence = 50\%:\\
\toprule
 0 & 1 & -0.0059 & 0.9941 & -0.59\% & 0.0200 & 0.0195 & 0.0205 \\
0.2085 & 1.2318 & 0.2079 & 1.2311 & -0.29\% & 0.0201 & 0.0205 & 0.0207 \\
 0.3551 & 1.4263 & 0.3551 & 1.4263 &0.00\% & 0.0202 & 0.0216 & 0.0212 \\
 0.4686 & 1.5978 & 0.4681 & 1.5970 & -0.11\% & 0.0203 & 0.0216 & 0.0216 \\
 0.5616 & 1.7535 & 0.5605 & 1.7515 & -0.20\% & 0.0204 & 0.0221 & 0.0220 \\
 \bottomrule
\end{tabular*}
\begin{tablenotes}\scriptsize
\item[$\dagger$ HR]: Hazard ratio
\item[$\ddagger$ ASE]: Average standard error
\item[$\dagger\dagger$ ESE]: Empirical standard error
\item[$\dagger\ddagger$ RSE]: Average robust standard error
\end{tablenotes} 
\end{table}
\end{center}

% Scenario 3----------------------

\begin{center}
\begin{table}[t]
\centering
\caption{Simulation results for time-varying treatment and covariates, sample size = 10,000}
\begin{tabular*}{500pt}{@{\extracolsep\fill}lccccccc@{\extracolsep\fill}}
\toprule
Prevalence = 25\%:\\
\toprule
True log  &  True marginal & Estimated log & Estimated & Avg & & & \\ 
 marginal  HR $\beta^{m_{2}}$& HR $e^{\beta^{m_{2}}}$  &  marginal HR $\overline{\hat{\beta}}^{m_{2}}$  &marginal HR $e^{\overline{\hat{\beta}}^{m_{2}}}$ & Bias &ASE &  ESE  & RSE \\
 \midrule
 0 & 1 & 0.0159 & 1.0160 & 1.60\% & 0.0232 & 0.0953 & 0.0827 \\
0.2085 & 1.2318 & 0.2230 & 1.2498 & 6.95\% & 0.0233 & 0.1049 & 0.0922 \\
 0.3551 & 1.4263 & 0.3701 & 1.4479 & 4.22\% & 0.0234 & 0.1235 & 0.0983\\
 0.4686 & 1.5978 & 0.4984 & 1.6460 & 6.36\% & 0.0235 & 0.1237& 0.0992\\
 0.5616 & 1.7535 & 0.5957 & 1.8142 & 6.07\% & 0.0236 & 0.1264 & 0.1025\\
\bottomrule
Prevalence = 50\%: \\
\toprule
 0 & 1 & 0.0050 & 1.0050 & 0.50\% & 0.0201 & 0.0475 & 0.0501 \\
0.2085 & 1.2318 & 0.2132 & 1.2376 & 2.25\% & 0.0201 & 0.0549 & 0.0547 \\
 0.3551 & 1.4263 & 0.3598 & 1.4330 & 1.32\% & 0.0203 & 0.0636 & 0.0586\\
 0.4686 & 1.5978 & 0.4779 & 1.6127 & 1.98\% & 0.0204 & 0.0606 & 0.0600\\
 0.5616 & 1.7535 & 0.5716 & 1.7711 & 1.78\% & 0.0206 & 0.0654 & 0.0629\\
\bottomrule
\end{tabular*}
\begin{tablenotes}\scriptsize
\item[$\dagger$ HR]: Hazard ratio
\item[$\ddagger$ ASE]: Average standard error
\item[$\dagger\dagger$ ESE]: Empirical standard error
\item[$\dagger\ddagger$ RSE]: Average robust standard error
\end{tablenotes} 
\end{table}
\end{center}

% Scenario 3 censoring -------------------------
\begin{center}
\begin{table}[t]
\centering
\caption{Simulation results for time-varying treatment and covariates with administrative censoring time $\tau =1 $,  sample size = 10,000}
\begin{tabular*}{500pt}{@{\extracolsep\fill}lccccccc@{\extracolsep\fill}}
\toprule
Prevalence = 25\%: \\
\toprule
True log  &  True marginal & Estimated log & Estimated & Avg & & & \\ 
 marginal  HR $\beta^{m_{2}}$& HR $e^{\beta^{m_{2}}}$  &  marginal HR $\overline{\hat{\beta}}^{m_{2}}$  &marginal HR $e^{\overline{\hat{\beta}}^{m_{2}}}$ & Bias &ASE &  ESE  & RSE \\
 \midrule
 0 & 1 & 0.0026 & 1.0026 & 0.26\% & 0.0418 & 0.0591 & 0.0669 \\
0.2085 & 1.2318 & 0.2396 & 1.2707 & 14.92\% & 0.0380 & 0.0603 & 0.0666 \\
 0.3551 & 1.4263 & 0.3938 & 1.4826 & 10.90\% & 0.0360 & 0.0612 & 0.0672\\
 0.4686 & 1.5978 & 0.5014 & 1.6510 & 7.00\% & 0.0347 & 0.0620 & 0.0685\\
  0.5616 & 1.7535 & 0.5904 & 1.8047 & 5.13\% & 0.0339 & 0.0646 & 0.0687\\
\bottomrule
Prevalence = 50\%: \\
\toprule
 0 & 1 & 0.0018 & 1.0018 & 0.18\% & 0.0363 & 0.0534 & 0.0614 \\
0.2085 & 1.2318 & 0.2334 & 1.2629 & 11.94\% & 0.0338& 0.0531 & 0.0598\\
 0.3551 & 1.4263 & 0.3861 & 1.4712 & 8.73\% & 0.0324 & 0.0493 & 0.0586\\
 0.4686 & 1.5978 & 0.4964& 1.6428 & 5.93\% & 0.0316 & 0.0506 & 0.0583\\
 0.5616 & 1.7535 & 0.5750 & 1.7771 & 2.39\% & 0.0311 & 0.0497 & 0.0573\\
\bottomrule
\end{tabular*}
\begin{tablenotes}\scriptsize
\item[$\dagger$ HR]: Hazard ratio
\item[$\ddagger$ ASE]: Average standard error
\item[$\dagger\dagger$ ESE]: Empirical standard error
\item[$\dagger\ddagger$ RSE]: Average robust standard error
\end{tablenotes} 
\end{table}
\end{center}

\begin{center}
\begin{table}[t]
\centering
\caption{Simulation results for time-varying treatment and covariates with administrative censoring time $\tau =0.25$, sample size = 10,000}
\begin{tabular*}{500pt}{@{\extracolsep\fill}lccccccc@{\extracolsep\fill}}
\toprule
Prevalence = 25\%:\\
\toprule
True log  &  True marginal & Estimated log & Estimated & Avg & & & \\ 
 marginal  HR $\beta^{m_{2}}$& HR $e^{\beta^{m_{2}}}$  &  marginal HR $\overline{\hat{\beta}}^{m_{2}}$  &marginal HR $e^{\overline{\hat{\beta}}^{m_{2}}}$ & Bias &ASE &  ESE  & RSE \\
 \midrule
 0 & 1 & 0.0032 & 1.0032& 0.32\% & 0.0984 & 0.1149 & 0.1201 \\
0.2085 & 1.2318 & 0.2972 & 1.3461 & 42.54\% & 0.0843 & 0.1054 & 0.1107 \\
 0.3551 & 1.4263 & 0.4891 & 1.6308 & 37.74\% & 0.0761 & 0.1023 & 0.1050\\
 0.4686 & 1.5978 & 0.6240 & 1.8664 & 33.16\% & 0.0705 & 0.0958 & 0.1007\\
  0.5616 & 1.7535 & 0.7333 & 2.0819 & 30.57\% & 0.0666 & 0.0945 & 0.0978\\
\bottomrule
Prevalence = 50\%: \\
\toprule
 0 & 1 & 0.0138 & 1.0139 & 1.38\% & 0.0855 & 0.1669 & 0.1612 \\
0.2085 & 1.2318 & 0.2944 & 1.3423 & 41.20\% & 0.0762& 0.1556 & 0.1529\\
 0.3551 & 1.4263 & 0.4947 & 1.6400 & 39.31\% & 0.0709 & 0.1419 & 0.1441\\
 0.4686 & 1.5978 & 0.6305& 1.8785 & 34.55\% & 0.0669 & 0.1371 & 0.1389\\
  0.5616 & 1.7535 & 0.7323 & 2.0799 & 30.40\% & 0.0639 & 0.1377& 0.1348\\
\bottomrule
\end{tabular*}
\begin{tablenotes}\scriptsize
\item[$\dagger$ HR]: Hazard ratio
\item[$\ddagger$ ASE]: Average standard error
\item[$\dagger\dagger$ ESE]: Empirical standard error
\item[$\dagger\ddagger$ RSE]: Average robust standard error
\end{tablenotes} 
\end{table}
\end{center}

From the estimating equation theory, when there is no censoring the IPTW estimate $\hat{\beta}^{m_{j}}$ converges to the log of the true log marginal hazard ratio $\beta^{m_{j}}$ as the sample size $n$ goes to infinity. Austin (2012)\cite{austin2} showed that without the presence of censoring, the IPTW estimator estimated the marginal hazard ratio for time to event data with negligible bias with sample size of 10,000. For our simulation studies with recurrent data, we obtained similar results. Table 2-4 summarize the results for scenarios without the presence of censoring. As we can see from the tables, the average bias is very close to 0 for independent gap time and time-varying covariates scenarios, while the bias is slightly higher for time-varying treatment and covariates scenarios. Hence it is safe to say that the IPTW estimator estimates the marginal hazard ratio for the second event with minimal bias when the sample size is 10,000. For simulation scenarios with 500 subjects, the bias is substantially higher compared to those with 10,000 subjects, especially for time-varying treatment and covariates scenarios. Thus, a sample size of 500 does not suffice to estimate the true marginal hazard ratio. Keeping sample size the same, the prevalence of treatment also impacts the degree of bias in estimating the marginal hazard ratio. We observe that the bias is lower when the treatment prevalence is 50\%. Table 5-6 summarize the results for scenarios with an administrative censoring time. With an administrative censoring time of 1, the IPTW estimator results in a moderate degree of bias in estimating the true marginal hazard ratio. Moreover, the bias tends to decrease as the true marginal hazard ratio is higher. As we increase the proportion of subjects being censored by setting $\tau$ to 0.25, the IPTW estimate tends to be further removed from the true marginal hazard ratio, regardless how large the sample size is. It is likely the estimate converges to another quantity other than the true marginal hazard ratio. However, it is not clear what the IPTW estimator is estimating in the presence of censoring. Therefore, we can conclude that the IPTW estimator results in biased estimation of the true marginal hazard ratio. For variance estimation, our results are in line with that of Austin's \cite{austin3}. Due to the within-subject correlation, the naïve variance estimator often tends to substantially underestimate the variance of the IPTW estimator\cite{Lin}. The robust variance estimator generally results in negligible bias in estimating the variance for independent gap time and time-varying covariate scenarios. The use of robust variance estimator results in minor bias for other scenarios, and the bias is generally within 15\%, which is still a huge improvement over the naïve variance estimator. 

\clearpage
%\begin{figure}[t]
%\centerline{\includegraphics[width=342pt,height=9pc,draft]{empty}}
%\caption{This is the sample figure caption.\label{fig1}}
%\end{figure}

%\begin{figure*}
%\centerline{\includegraphics[width=342pt,height=9pc,draft]{empty}}
%\caption{This is the sample figure caption.\label{fig2}}
%\end{figure*}
\clearpage
\section{Discussion}\label{sec3}
We conducted an extensive series of Monte Carlo simulations to examine the performance of the inverse probability of treatment weighting (IPTW) method for estimating the marginal hazard ratio in the context of recurrent events. We briefly summarize our simulation results and discuss what can be done in subsequent research.

Previous research has shown that the inverse probability of treatment weighting (IPTW) method results in unbiased estimation of the marginal hazard ratio for time to event data \cite{austin2}. We made a step further to investigate the performance of the IPTW method for estimating marginal hazard ratio for recurrent data. We found that without the presence of censoring, the IPTW estimator consistently estimated the marginal hazard ratio for the first and second event across all scenarios. However, sample size and treatment prevalence both had an impact on the accuracy of the estimation of the marginal hazard ratio. Though the IPTW estimator is asymptotically unbiased, the estimate could be biased for studies with small sample size. Therefore, we recommend researchers use the IPTW estimator to estimate the marginal hazard ratio only when the sample size is sufficiently large. In the presence of censoring, we observed from the simulation results that the IPTW estimator resulted in biased estimation of the marginal hazard ratio. The degree of bias tended to be greater as we increased the proportion of subjects being censored. The estimate seemed to converge to another quantity, however, we are unsure of what it converges to and the interpretations of the estimate remains unclear. Variance estimattion plays an important role in determining the optimal sample size if one would like to control the average bias of the IPTW estimate to be within a certain range. Austin (2016) showed that the use of the robust variance estimator significantly improved accuracy of variance estimation \cite{austin3}. Based on our simulation results, it turned out that the robust variance estimator approximated the variance reasonably well within 10\% bias for most scenarios, while the naïve variance estimator often substantially underestimated the variance. Due to time constraint, we did not consider the use of bootstrap variance estimator, which was proven to have similar performance to that of the robust variance estimator for time to event data \cite{austin2}.

Certainly there are some limitations in our simulation studies. One of the limitations is that we only considered one dependence relationship between $X(1)$ and $X(2)$. Though the IPTW estimator consistently estimates the marginal hazard ratio for recurrent data, the converging rate may differ for different dependence relationship between $X(1)$ and $X(2)$. If the researcher would like to control the estimation bias within a certain range for a study, it would be difficult to figure out the optimal sample size. Another limitation is that we only considered the case where $X(1)$ follows a standard normal distribution. Further research can be done to investigate the behaviour of the IPTW estimator when $X(1)$ is binary or categorical. We could also incorporate more variables or even consider the possibility of dependence relationship among these variables. One could also consider other propensity score methods such as matching, stratification to estimate the marginal hazard ratio.

To summarize, based on the simulation results, we recommend researchers use the IPTW estimator to estimate the marginal hazard ratio with the robust variance estimator for observational recurrent data without the presence of censoring. However, researchers should be aware that though the IPTW estimator is asymptotically unbiased, when dealing with small samples the IPTW estimator may result in bias in estimating the marginal hazard ratio. To this end, extra efforts should be paid to determine the minimal sample size. In the presence of censoring, the IPTW method results in biased estimation of the marginal hazard ratio, and the degree of bias is related to the proportion of subjects being censored. For this reason, the IPTW estimator should not be used in the presence of censoring. Subsequent research should investigate methods for estimating the marginal hazard ratio or figure out what the IPTW estimator estimates in the presence of censoring. Further extensions can be made to settings with multiple events and the stabilized weights can be formulated in the similar way to that described in Section 2.1 \cite{Hernan}.

\nocite{*}% Show all bib entries - both cited and uncited; comment this line to view only cited bib entries;
\bibliography{wileyNJD-AMA}%

\clearpage

\end{document}